# Enhanced Voice Post Processing Using Voice Decoder Guidance Indicators

Phani Kumar Nyshadham, D R Shivakumar,  Peter Kroon, and Shmulik Markovich-Golan

Intel Corporation, Santa Clara, USA

## ABSTRACT

Voice enhancement and voice coding are imperative and important functions in a voice-communication system. However, both functions are commonly treated independently, even though both utilize similar features of the underlying signals. Our proposal is to leverage information from one function to the benefit of the other. Specifically, our proposed changes are focused on changes to the voice enhancement at the downlink side and utilizing information of the voice decoding. Preliminary results show that such an approach results in improved quality. Additionally, suggestions are provided on future extensions of the proposed concept.

## 1. INTRODUCTION

In digital voice communications (e.g., in voice or video calls), the quality of the voice signal as perceived by the end user is one of the most important measures which determine the efficiency of the communication system.  The quality of the signal is based on many factors, but for the receiving end it will be mostly determined by the quality of the decoded signal and any subsequent post processing

During a voice call, the voice signal suffers from interference of undesired signals and noise, as well as distortions. Several interferences like echoes, different types of stationary/non-stationary noises like ambient and environmental noises contaminate the desired signal. Voice pre-processing or enhancement at the transmitter side typically enhance the desired signal by suppressing the various interferences, however, for strong interferences, the desired signal might be distorted in the process. Furthermore, during transmission additional distortions are added (resulting from level differences, network noise, etc.).  The downlink post-processing or enhancement aims at reducing the artifacts of the received voice signal, thus improving the quality of the underlying voice signal at the receiver. A simplified block-diagram of a voice communication system is depicted in Figure 1.

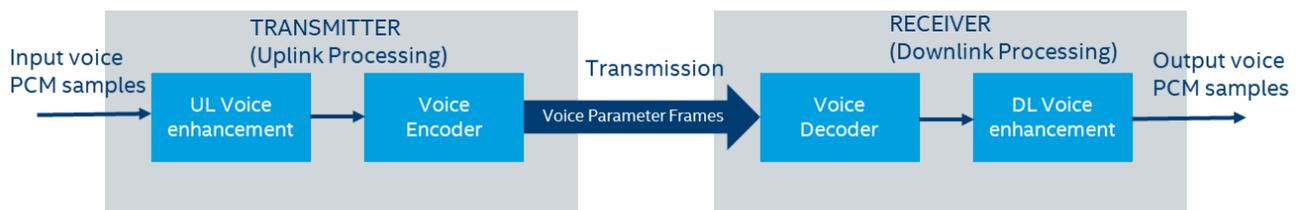

*Figure 1: Simplified block-diagram of a voice-communication system.*

Voice enhancement algorithms are mostly based on statistical methods which utilize estimated features of the noisy and distorted voice signal. Common signal features include noise spectrum, signal spectrum, spatial covariance matrix, steering-vector, speech-presence-probability, speech-activity and dynamic-range. These features are then utilized for enhancing the signal, either by using simple physical models for the signals or by using hidden and more complicated relations between the signal and its features that are derived using deep-learning methods.

In this paper, we deal with this problem of reliability limitations of signal characterization parameters faced by the downlink voice enhancement algorithms which typically operate in a stand-alone unguided manner using only the decoded voice signal as its input [1]. In [2] a post-filter for reducing voice-coding artifacts using decoder information was proposed. However, this proposal does not consider the more generic voice enhancement problem, trying to improve voice-quality of a voice signal contaminated by various noise, interference and distortion, by leveraging information from the voice-coder.
We propose a method to improve the reliability of signal characterization parameters, thereby improving the performance of downlink voice enhancement algorithms.

Another disadvantage of applying voice-enhancement and voice-coding independently is an unavoidable tradeoff between voice-quality, latency and computational power. Modern voice-encoders construct advanced models of the voice signal, thereby analyzing it extensively and accurately. Obtaining such models require sufficient observation time (typically 20ms) and computational resources. This information, though available, is not used by the voice-enhancement at the receiver side. Instead, either simple signal processing, which prefer low latency and low complexity over voice-quality and only compensate for the loudspeaker limitations and properties, is applied, or more advanced voice–enhancement techniques, such as noise-reduction, which prioritize voice-quality over latency and compute power, is applied.

In this paper, we propose to improve the voice quality by leveraging information obtained from the voice decoder and incorporating it in the downlink voice-enhancement, also denoted as *guided voice-enhancement*. This is contrary to existing voice-communication which treat voice-coding and voice-enhancement separately and independently, denoted as unguided voice-enhancement. The information that can be leveraged by the downlink voice-enhancement is derived from the coded bitstream and decoder state information including voice-signal model classification and parameters, noise classification, noise spectrum and speech activity.

A first example of the proposed method is utilizing the voice parameters contained in the transmitted voice frames, see Figure 2.

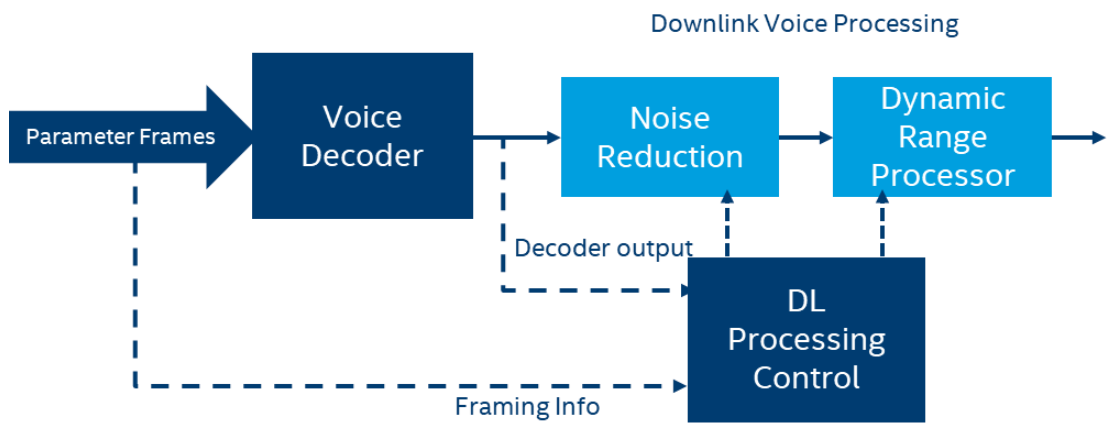

Figure 2: Frame guided downlink voice-enhancement.

This is referred to as *Frame guidance*, and an example will be explained in detail below.

Although this approach will have its immediate benefits, not all signal features are readily observable from the voice parameters.  This is due to the algorithmic principles used by most voice coders, where prior analyzed (and decoded) signals are used to reconstruct signals for the current frame.  Hence, in the second example, depicted in Figure 3, more detailed insights of the underlying signal features can be obtained from the internal decoder states.  Although the principles of using these decoder state variables will be similar, the specific details will vary for the different voice coders used.

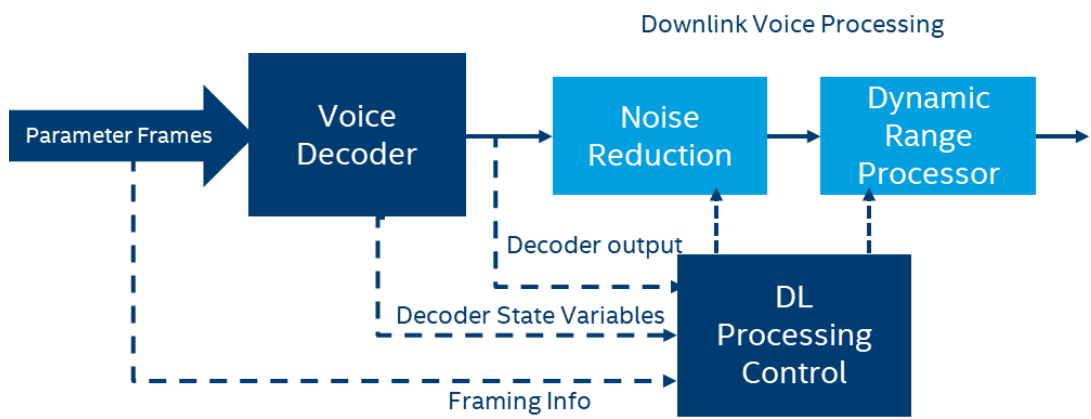

*Figure 3: Frame and decoder guided downlink voice-enhancement.*

This approach is referred to as *Frame and Decoder Guidance*.  Of course it is also possible to only use the Decoder Guidance, but most benefit can be achieved if all available information will be used.

Some of the features of the voice-decoder can be readily incorporated in statistical model-based voice-enhancements, whereas other relations between the features of the encoded noisy signal and the clean voice-signal are more intricate, and require applying deep-neural network methods for revealing and utilizing them. In the following sections, we show some examples of improving the voice quality by leveraging information of the voice-decoder in the downlink voice enhancement. Specifically, leveraging noise-spectrum and voice-signal classification for further reducing noise, both stationary and non-stationary, detecting background music and preserving it and leveraging voice-activity-detection for controlling the dynamic-range conversion curve of the loud-speaker signal.

## 2. PROPOSED SOLUTION

Voice encoders and decoders are standard components in digital voice communication systems. They use sophisticated algorithms to reduce the amount of information to describe the voice signal.  This compressed signal is transmitted as parameters to the receiver and used by the decoder to render a perceptually similar sounding version of the original voice signal.  The conventional solution using both voice decoder and voice enhancement is shown below, in Figure 4, which we refer to as unguided voice enhancement. The enhancement modules shown are just for illustration, and additional modules could be used.

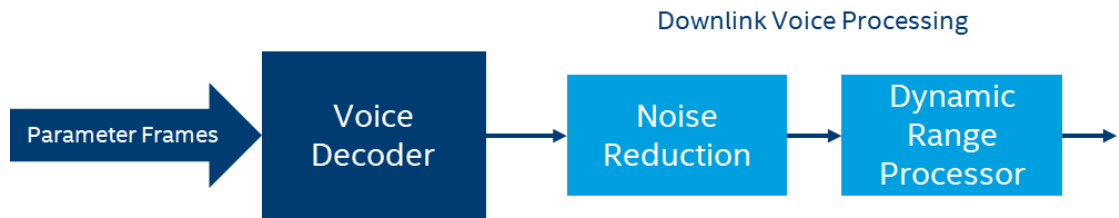
*Figure 4: Conventional downlink processing with unguided downlink voice-enhancement.*

One common approach used by most coders in cellular communication systems, is to only transmit voice parameters during active voice, and send a lower rate information stream during non-active voice. This principle is referred to as DTX (Discontinuous Transmission) [4], and the lower rate information is referred to as Silence Description or SID frames.   Note that SID frames can describe not only silence but also noise with different levels and characteristics. To enable this feature the voice encoder needs to perform Voice Activity Detection [5] on the input signal, and the outcome of this classification is explicitly transmitted in the voice parameters.

In the examples below, we consider the EVS [3] voice-coder because it is the state-of-the-art coder which is used for Voice-over-LTE use-cases for 4G/5G Radio Access Technologies. It is capable of reducing the communication bandwidth without sacrificing much signal quality by classifying the signal type and obtaining accurate models for it. In turn, the richness of the extracted parameters make it a good example for showing the advantages of the proposed idea. Nevertheless, the proposed techniques will also benefit other commonly used voice coders such as AMR-NB and AMR-WB coders.

The rest of this section is structured as follows. In Section 2.1 we elaborate about the information that can be extracted from the voice-decoder.  Later in Section 2.2 we specify how that information can be leveraged in the voice-enhancement for improving voice quality. Finally, in Section 3 we show how the proposed idea can be used for improving voice quality in some specific examples.

### 2.1. Voice-coder information

We distinguish between two guidance levels, namely frame guidance and decoder guidance.

*Using Frame Guidance*

The voice decoder uses the information stored in the encoded voice frame and provides voice samples in PCM format as output.  The decoder operation starts by analyzing of header of the encoded frame.  A frame structure encoded by the EVS coder according to 3GPP standards for 4G/5G RATs is depicted in Figure 5.

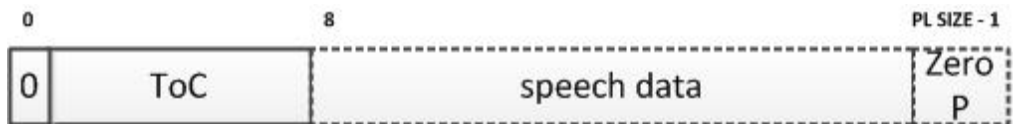
*Figure 5: EVS frame structure.*

The structure of the Table of Content (ToC) field of the header is depicted in Figure 6.

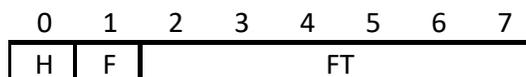
*Figure 6: ToC field structure.*

The Frame Type index (denoted as FT), can be set to either *EVS Primary*, *EVS AMR-WB IO*, or *comfort noise* (SID). The FT field also contains: EVS mode (1 bit), Unused/Q bit (1 bit) depending on the value of EVS mode bit, and EVS bit-rate (4 bits).

Therefore, from examining the ToC field of the encoded frame, it can be categorized as

1. SPEECH frame
2. SID frame (noise)
3. SPEECH_LOST
4. NO_DATA

From the above 4 frame-types, SPEECH frame implies that the encoded frame is actually a SPEECH frame. SID and NO_DATA imply that the encoded frame contains no voice activity and is described by SILENCE descriptor. SPEECH_LOST is kind of BAD FRAME INDICATOR.

Therefore, the EVS frame header can be used for voice-activity-detection.

### Using Decoder Guidance

Several internal states of the voice decoder contain signal features that can be leveraged by voice enhancement modules. For example:

*Noise features:* SID frames contain information about the noise spectrum, to be used for comfort-noise-generation (CNG). The noise spectrum can be used by the noise-reduction in the voice-enhancement, thereby reducing its computational complexity. Typically, the noise spectrum is estimated based on the minimum-statistics principle.

*Voice pitch and linear-prediction coefficients:* These parameters are key to multiple voice-enhancement operations such as bandwidth extension, lost-packets concealment and slowing down voice. Extracting pitch and linear prediction coefficient is compute intensive, and leveraging precomputed values will reduce the computational complexity of the voice-enhancement. The signal model can be also utilized for noise-reduction for improving voice quality.

*Coding mode detection:* The decoder has capability to detect the coding mode which can differentiate between voice and music signals. This is a very important state information that can control the degree of noise reduction based on the signal classification.

Several other decoder states not limited to the above examples can be used for the mechanism stated in this paper. We propose to feed the FRAMETYPE and DECODER STATES which convey very reliable information about objective and subjective signal features to the voice enhancement chain. This technique is of course applicable to other voice coders such as AMR-NB, AMR-WB, EVS, EVS-WB-IO, etc.

## 2.2. Leveraging voice-decoder information in downlink voice-enhancement

The extracted information of the voice-decoder stage can be leveraged by the downlink voice-enhancement in different methods. A straight forward method, denoted as model-based controlling downlink voice-enhancement and depicted in Figure 7, is to compute some of voice-enhancement state variables as a function of the framing information, the decoder state variables and the decoder output. Noise reduction and Dynamic Range Processor are examples of downlink voice enhancement modules, and of course others can be incorporated. Although simple, this method is sub-optimal, as it is not exploiting more intricate relations to the voice-coder features. Two additional methods, incorporating deep-neural-network methods are denoted as DNN-based controlling downlink voice-enhancement and DNN-based downlink voice-enhancement are respectively depicted in Figure 8 and Figure 9. The DNN-based controlling downlink voice-enhancement method aims to reveal hidden and non-trivial relations between voice-enhancement and voice-coder state variables. The DNN-based downlink voice-enhancement method combines both the control and voice-enhancement functions and aims to find a direct relation between the framing information, the decoder state variables and the decoder output and between the enhanced desired voice signal.

To emphasize the difference between the various methods let us consider a specific case where the voice linear-prediction coefficients, obtained from the voice-decoder, are incorporated in the noise-reduction module in the downlink voice-enhancement. In the model-based controlling downlink voice-enhancement we would replace the noisy-signal spectrum estimate with a computation derived from the LPC model and frequency domain (FD) SID model. In the DNN-based controlling downlink voice-enhancement method, we could additionally train a DNN to estimate the speech-presence-probability (per frequency) from the noisy LPC coefficients and FD-SID estimates of the noisy signal. And finally, in the DNN-based downlink-processing, we could train a DNN to estimate the spectrogram of the clean signal from the noisy LPC coefficients.

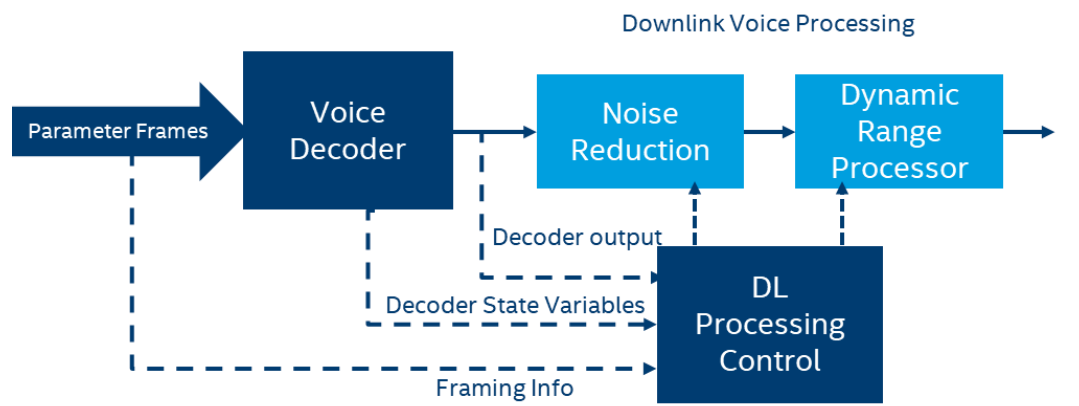

*Figure 7: Model-based control of downlink processing.*

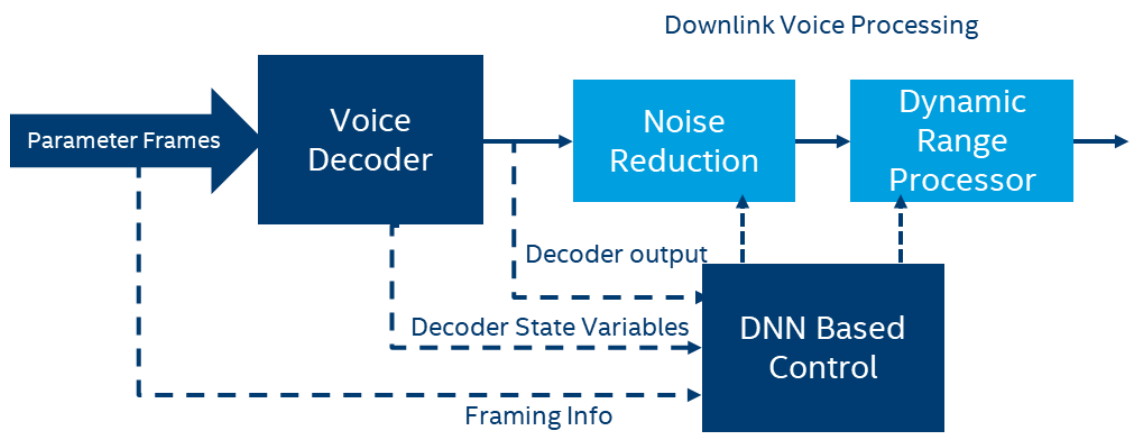

*Figure 8: DNN-based control of downlink processing*

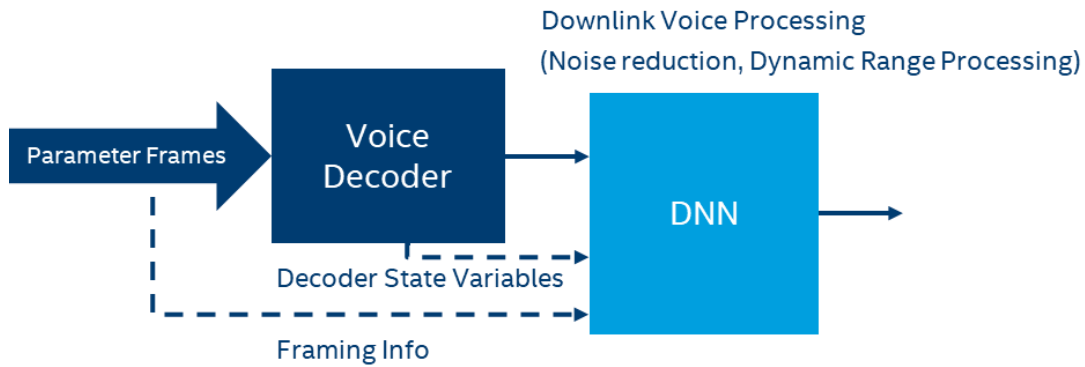

*Figure 9: DNN-based downlink voice-enhancement.*

Aside from performance improvement, the decoder guidance can contribute to reducing the computational complexity of the voice-enhancement, thereby saving power, by reusing some of its features and by reverting the voice-enhancement to an idle state (where applicable) during silence frames. In other cases, utilizing signal classification, such as music, can trigger on and off compute intensive operations.

## 3. PRELIMINARY EXPERIMENTAL RESULTS

In this section we provide details of various cases in which the proposed idea is utilized. Specifically, we consider:
- Improving noise reduction for various noise conditions (see in the following NR section).
- Music signal preservation while performing noise reduction
- Improving dynamic-range of the loud-speaker signal (see in the following MDRP section)
- Additional applications

### 3.1. Noise Reduction (NR)

We compare the performance of an unguided NR and a guided NR. Single-channel noise-reduction methods, e.g., Wiener filter, spectral-amplitude (SA), log SA (LSA), optimally-modified LSA (OMLSA), etc., rely on estimated noise power-spectral-density (PSD) and instantaneous signal-to-noise-ratio (SNR) estimates to compute and apply a time-varying spectral gain which optimizes a pre-defined performance criterion.

The noise-PSD stage is typically comprised of either: a) a voice-activity-detector (VAD) that enables the PSD tracking of the noisy-speech signal during noise-only time-segments; or b) a minimum-statistics estimate of the noise PSD (assuming that speech and noise components are statistically-independent, the spectrum of the noise-speech signal equals the sum of their individual spectra, and further assuming that the noise spectrum is slowly time-varying, the minimum of the time-varying PSD estimate is proportional to the noise spectrum).

The computational complexity of both VAD estimation and minimum-statistics based noise-PSD estimation can be substantially reduced by leveraging the encoded information in the voice-coder packets, specifically, silent frame indication.

The proposed guided NR using a model-based control is depicted in Figure 10.
The block-diagram of the noise-reduction is simplified and comprises of the following components:
1) Noise Power estimation: The signal is converted into frequency domain using spectral analysis. The power of signal is computed for every frequency bin. The power of the signal is then recursively averaged using minimum-statistics estimator which produces smoothed noise estimates for every frame.
2) Suppression gain computation: The smoothed noise estimates are used for by a gain computation module, followed by spectral subtraction for reducing the noise power. The gains are computed for every frequency bin after which the frequency domain signal is converted to the time domain.

The FRAMETYPE and DECODER STATE variables (noise power estimates) are fed to a new DL processing controller, comprising:
1) Signal Classifier: examines the signal characterization parameters (as part of decoder states) and classifies the signal as either VOICE or MUSIC signal. If the SIGNAL MODE is music, DL processing control BYPASSES the NR module. If the SIGNAL MODE is voice, the control flow proceeds to FrameType detector.
2) Frame type detector: If the FrameType is detected as INACTIVE (SID/NO_DATA), the decoder states for noise estimates are passed to the suppression gain computation. If the FrameType is ACTIVE SPEECH, NR noise estimates are passed to the gain computation of NR module. The Noise estimate multiplexer chooses NR estimates or DECODER STATE estimates based on FRAMETYPE.

This way, reliable noise estimates are used for gain computation in case of INACTIVE frames, thus improving performance.

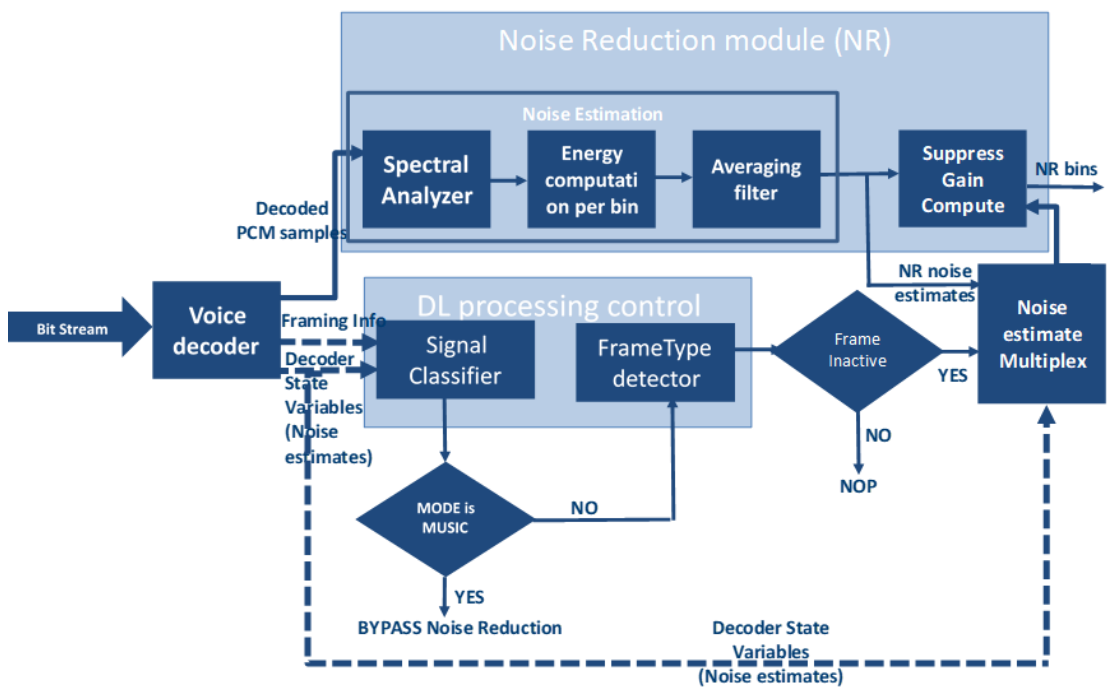

Figure 10: Model-based controlling of downlink noise-reduction.

Another variation of the implementation could be to stop the NR estimation block of NR module and use the decoder state variables for noise estimation instead for INACTIVE frames. Then NR state variables are initialized with the DECODER STATE variables for seamless execution. This way, both the computational complexity of NR module is reduced for INACTIVE frames and the performance is improved.

The performance of the proposed guided downlink noise reduction is evaluated in various noise conditions (white, train, road, pink, car, cafeteria and wind) and compared to the performance of an unguided noise reduction. The results are summarized in Table 1. Evidently, guided noise-reduction outperforms unguided noise-reduction, by 3.7dB-10.1dB, depending on the noise type. For the case of white noise and wind noise we depict the noise-reduction as a function of time in Figure 11 and Figure 12, respectively. Clearly from these figures, the advantage of guided NR over unguided NR is consistent.

| Type of Noisy signal | Noise Suppression Unguided NR [dB] | Noise Suppression Guided NR [dB] | Improvement [dB] |
|---|---|---|---|
| White Noise | 15.47 | 19.21 | 3.74 |
| Train noise | 4.36 | 10.15 | 5.79 |
| Road noise | 10.58 | 17.57 | 6.99 |
| Pink Noise | 9.54 | 15.2 | 5.66 |
| Crossroad | 2.88 | 10.2 | 7.32 |
| Car | 2.05 | 12.12 | 10.07 |
| Cafeteria | 1.71 | 6.28 | 4.57 |
| Wind | 0 | 9.98 | 9.98 |

Table 1: Performance comparison between proposed guided noise-reduction and unguided noise-reduction for various noises.

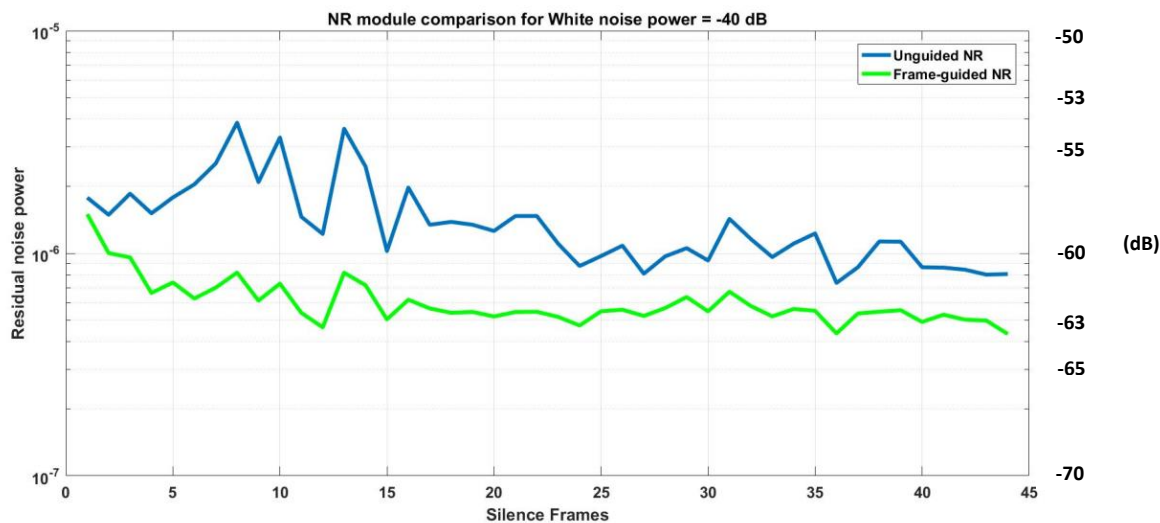

Figure 11: Residual white noise power for guided NR (in green) and unguided NR (in blue) as a function of time.

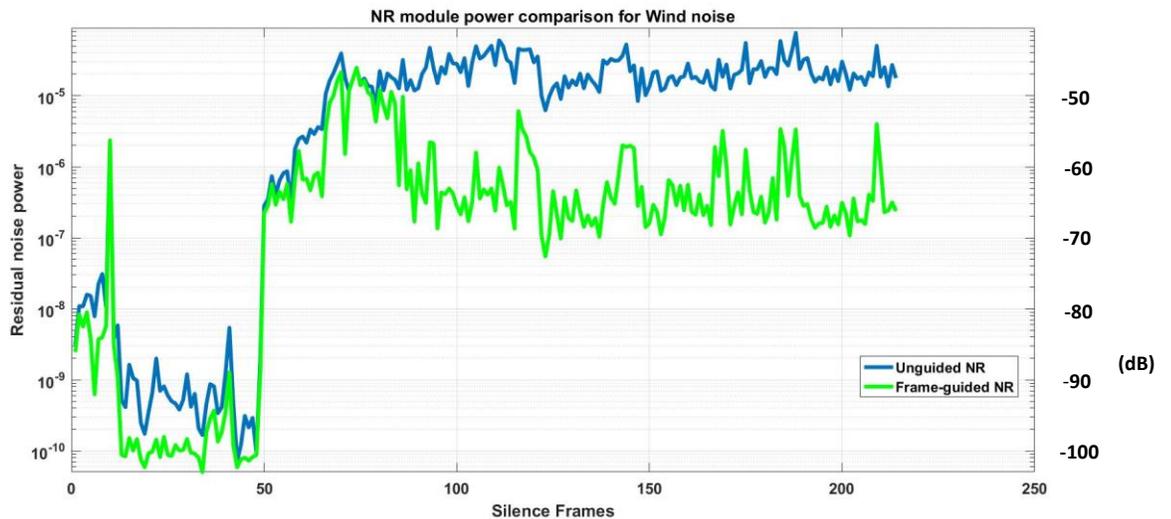

*Figure 12: Residual wind noise power for guided NR (in green) and unguided NR (in blue) as a function of time.*

*Music signal preservation while performing Noise Reduction*

In the following example, we evaluate the performance of NR module with an input signal comprising of noisy speech and background music. The time-domain signals of the input, and outputs of unguided NR and guided NR, and their spectrograms are depicted in Figure 13, where the yellow squares mark time-segments contain music and unmarked time-segment contain speech. Conventional NR which operates in unguided manner applies the same level of noise reduction to the whole signal (see Figure 13-(*b*), and as a result leading to music distortion. The proposed guided NR uses the signal classification parameter defining speech/music that is obtained from the decoder state variables. In The guided NR module knows with high reliability whether the signal is speech or music, and can adjust its function accordingly. During speech frames the guided-NR performs a more aggressive noise reduction improving the quality of speech and during music frames it applies a moderate and soft noise reduction, thereby preserving the music content and avoiding severe spectral distortion deteriorating its harmonic structure (see Figure 13-(*c*).

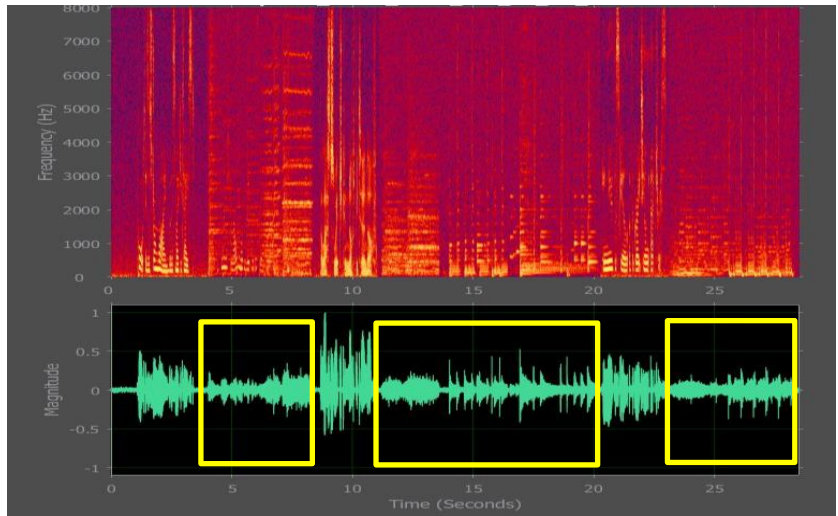

*(a) Microphone signal*

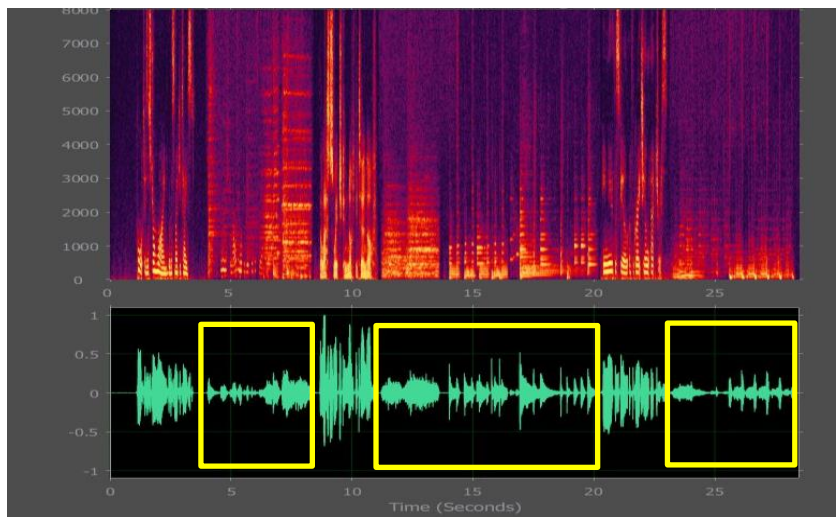

*(b) Unguided noise-reduction.*

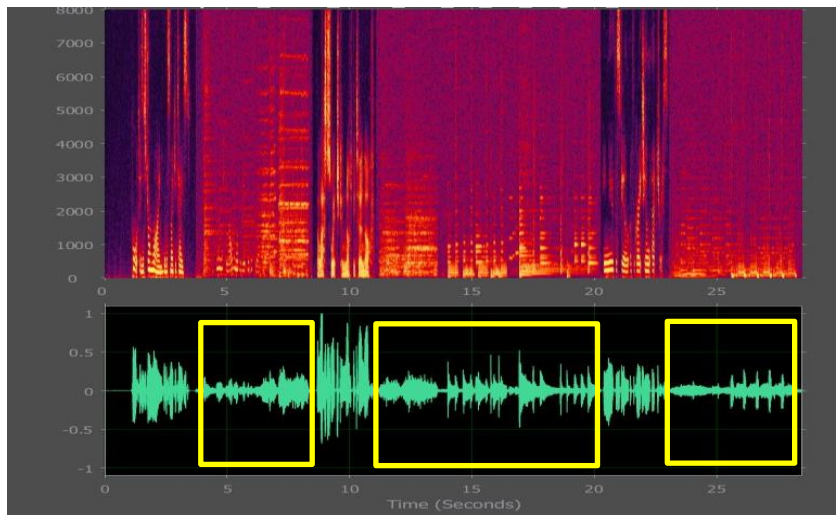

*(c) Guided noise-reduction*

*Figure 13: Noise reduction of noisy speech with background music: Input signal (Left), Unguided NR (Right)*

## 3.2. Multi-Band Dynamic Range Processor (MDRP)

As another example of voice enhancement module, we analyzed MDRP which improves the dynamic range of the downlink decoded voice signal for better listening experience. Multiband Dynamic Range Processor (MDRP) performs an audio compression/expansion on multiple sub-bands of the voice signal. For each sub-band, MDRP applies piece-wise linear gain to the signal based on the power thresholds. This compression/expansion ensures rich listening experience by improving the contrast of the voice signal.

The input and output signals of the unguided MDRP and guided MDRP and their spectrograms, are respectively depicted in Figure 14 and Figure 15. The yellow squares in the figures mark noise-only time-segments. As can be seen in the unguided MDRP case in Figure 14, it boosts up the noise trying to improve its dynamic-range while assuming it is desired signal. The guided MDRP, however, uses the FrameType information obtained from the decoder state for identifying SILENCE time segments, during which it reverts to a different dynamic-range curve, avoiding undesired noise boost, see in Figure 15. Note, that due to a minimum silence period at the voice-encoder, there is an offset in the SILENCE indication resulting in a temporary boost of at the beginning of noise-only segments. Of course, every algorithm can be tuned to smoothen the transition between segments having different degrees of enhancement with decoder state guidance by interpolation techniques.

Aside from the clear quality advantage, the guided MDRP can also contribute to computational complexity saving, if it operation is gated by and enabled only during speech frames.

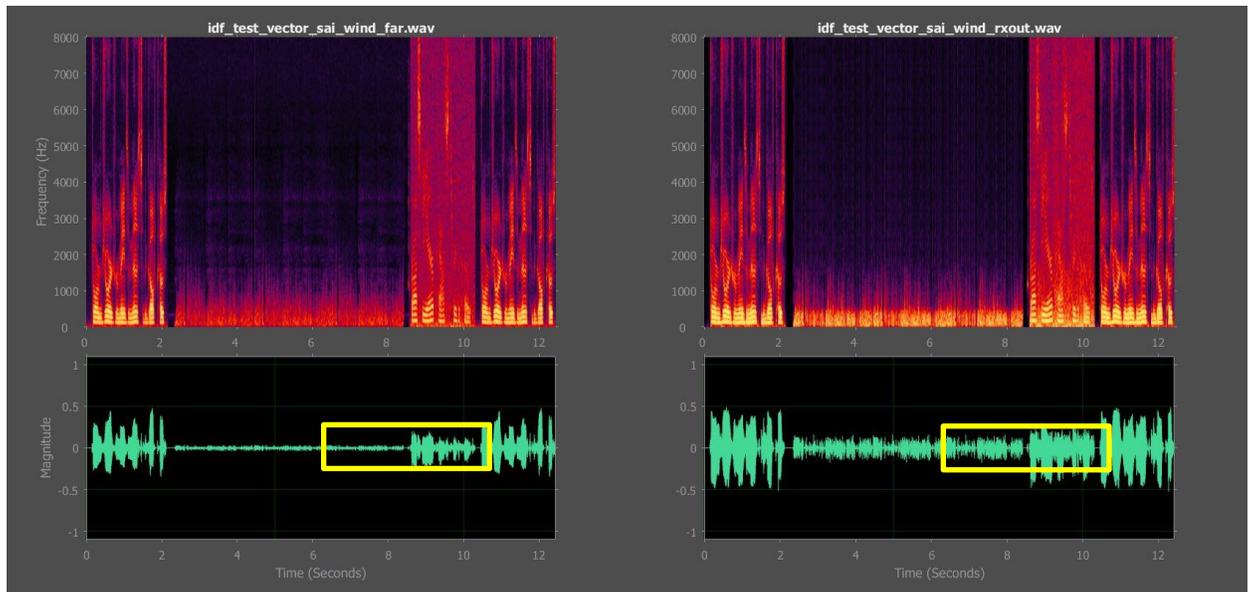

*Figure 14: Input signal (left) and output signal (right) of an unguided MDRP.*

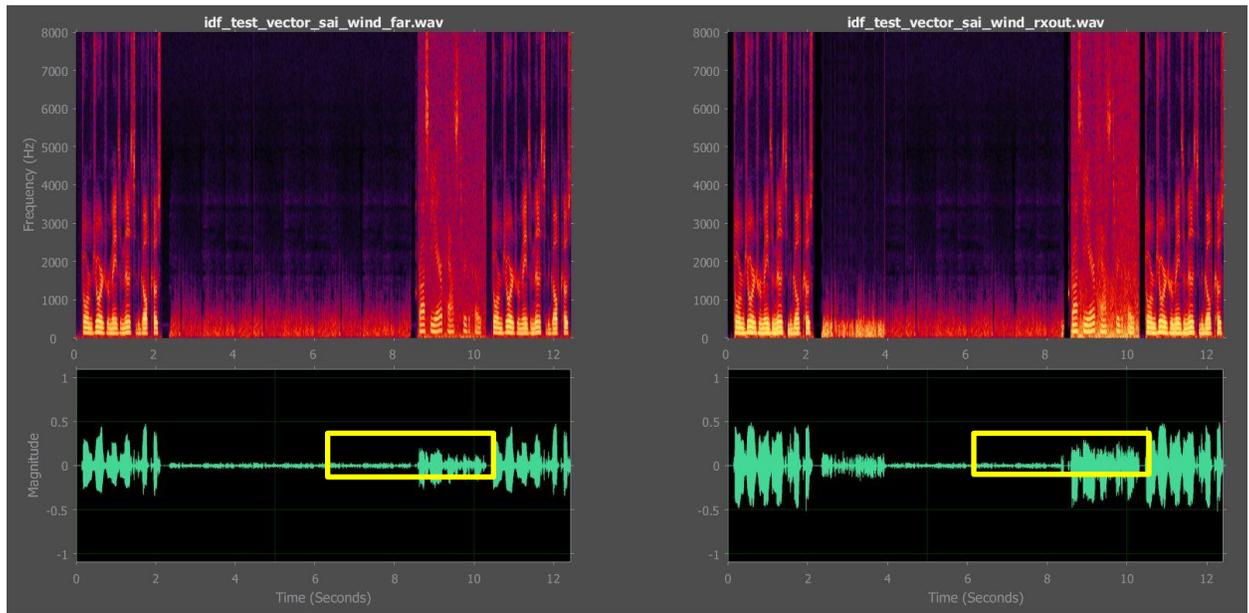
*Figure 15: Input signal (left) and output signal (right) of a guided MDRP*

### 3.3. Additional applications

Other examples of voice-enhancement modules that can benefit from the silence indication of the decoder are:

*Acoustic Echo Cancellation (AEC):*

An uplink voice enhancement module which is responsible for cancelling the echo signal (downlink signal (DL)) leaked from the speaker into the microphone. The algorithm starts by detecting if there is any voice activity in the leaking DL signal components. If there is no activity, the algorithm should stop adapting to avoid its divergence.

*Wind Noise Reduction (WNR):*

A downlink voice enhancement module which is responsible for reducing wind noise. We measured the silence-frame detection rate with and without guidance for two test vectors. In both vectors, decoder guidance had a significant impact on the silence detection rate, boosting it from 52.4% and 7.6% to 100%. This has two desired impacts. First, voice-quality is improved, since the noise reduction is more consistent, and mistakenly switching between noise frames to speech frames is avoided. And second, the computational complexity is significantly reduced, as the number of computations is significantly lower during noise-only time-segments.

## 4. CONCLUSION

This paper proposed enhancing downlink voice enhancement by leveraging the information provided by the decoder. The advantages of the proposed idea is threefold, improving the quality of the voice signal without increasing latency or computational complexity. Preliminary results were provided showing that the quality of the voice signal is improved, where we quantify it by SNR improvement in various noise conditions, preservation of music background and dynamic-range. This is done without increasing computational complexity or latency, by leveraging previously estimated signal features. Explicitly, the computational complexity and required latency have already been invested in the voice-encoder in the uplink side, or in the downlink-decoder in the downlink side.

Note that current solutions, in which voice-enhancement is performed independently of voice-coding, are inferior due to computational complexity and latency constraints, which are alleviated in the proposed idea.

## 5. REFERENCES

[1] Rabiner, Lawrence R. "Applications of voice processing to telecommunications." Proceedings of the IEEE 82.2 (1994): 199-228.

[2] Chen, Juin-Hwey, and Allen Gersho, "Adaptive postfiltering for quality enhancement of coded speech." IEEE Transactions on Speech and Audio Processing 3.1 (1995): 59-71)

[3] M. Dietz et al. Overview Of The EVS Codec Architecture, Proceedings ICASSP, 2015

[4] 3GPP TS 26.450 , Codec for Enhanced Voice Services (EVS); Discontinuous Transmission (DTX), 2015

[5] 3GPP TS 26.451, Codec for Enhanced Voice Services (EVS); Voice Activity Detection (VAD), 2015